# udpPacketManager: An International LOFAR Station Data (Pre-)Processor

David J. McKenna [1, 2], Evan F. Keane [2], Peter T. Gallagher [1], and Joe McCauley [2]

**1** Dublin Institute for Advanced Studies, Ireland **2** Trinity College Dublin, Ireland

## Summary

International LOFAR stations are powerful radio telescopes, however they are delivered without the tooling nessasary to convert their raw data stream into standard data formats that can be used by common processing pipelines, or science-ready data products.

udpPacketManager is a `C` and `C++` library that was developed with the intent of providing a faster-than-realtime software package for converting raw data into arbitrary data formats based on the needs of observers working with the Irish LOFAR station (I-LOFAR), and stations across Europe. It currently offers an open-source solution for both offline and online (pre-)processing of telescope data into a wide variety of formats.

## Statement of need

International LOFAR stations (van Haarlem et al. 2013) produce a 3 Gbps stream of UDP packets, split across 4 separate ports. Each packet contains a standard header with basic time and telescope hardware information, typically followed by 7808 bytes of time-major beamformed voltage data, with 2 complex samples per antenna polarisation, for 16 time samples across a variable number of frequency samples (Virtanen 2018, Table 6). In order to convert these data into a usable format, the 4 ports must be combined, data integrity issues (such as packet loss) must be identified and mitigated, and the beamformed samples must be unpacked and reordered following a set specification in order to be processed efficiently.

Previous work in this regard includes the LOFAR und MPIfR Pulsare (`LuMP`) Software (Anderson 2013), an open-source recorder that saves data to a PUMA2-derived data format that can be parsed using the DSPSR (W. van Straten and Bailes 2011) ecosystem, and the ARTEMIS system (Serylak et al. 2012), a hardware-software package for online transient observations derived from the source-available `PELICAN LOFAR` backend (Developers 2021). These were found not to be sufficiently flexible to account for peculiar ways of utilising the telescope hardware, such as supporting multi-mode observations (McKay-Bukowski 2013), nor work within some constraints of the REALTA compute cluster (Murphy et al. 2021). Consequently, `udpPacketManager` was built to better facilitate observations with the telescope.

The software has supported ongoing observations of the Sun, Pulsars and Rotating Radio Transients (Murphy et al. 2021; McKenna et al. 2023) in Ireland since early 2020, alongside multi-site work with Breakthrough Listen in the search for extraterrestrial life (SETI) in coordination with the Sweedish LOFAR station at Onsala since 2021 (Johnson et al. in prep.) and observations of Jupiter in coordination with the French LOFAR station and one of the German LOFAR stations, at Postdam, in 2022 (Louis et al. 2022).

## Inputs, Outputs and Metadata

`udpPacketManager` contains a flexible interface for handling multiple different data sources and sinks in a semi-transparent manner, allowing for data to be read and written out through normal files, FIFO named pipes, `zstandard` (Collet and Kucherawy 2021) compressed files (which allow for a greater than two times space-saving when handling raw voltages from LOFAR stations), `PSRDADA` ringbuffers for online processing (Willem van Straten, Jameson, and OsAowski 2021), and write-only to `HDF5` files that conform to LOFAR ICD003 (Alexov et al. 2012). These interfaces were also used to create the associated `ILTDada` project (McKenna 2023), a UDP packet capture software that writes data to a `PSRDADA` ringbuffer for immediate processing with this software.

Given that the header attached to the data itself contains minimal metadata outside the start time of the data block, metadata is parsed from an associated file that contains one or more commands to control the telescope (`beamctl` (Virtanen 2018, Appendix A.2)) and minimal other metadata, such as observer and project identifiers. This is then processed to generate a chosen output header, which can follow the `GUPPI RAW` (Lebofsky et al. 2019), `SIGPROC` (Lorimer 2011), `PSRDADA`, or ICD003 specifications.

The library can reform the data to a number of output formats. For raw voltages outputs, standard time-major and beamlet-major orders, which can also be additionally split by antenna polarisation and complex sample components are available. Partial or full Stokes parameters are also generated in similar formats, with temporal downsampling up to a factor of 16 available in the library itself. All of these outputs can be calibrated to correct for interferometric beam issues by applying, or generating for external use, Jones matrices generated by `dreamBeam` (Carozzi 2023).

Two command-line-interfaces are provided alongside the library. The first, `lofar_udp_extractor`, offers an easy interface for accessing the normal functionality of the library for observers, while the second, `lofar_stokes_extractor`, utilises the output voltages to perform additional processing for science-ready outputs, such as channelisation of Stokes parameters through the `FFTW` library (Frigo and Johnson 2005) and additional temporal downsampling beyond the library's factor of 16 limit. Both of these interfaces act further act as examples of how to further utilise the library and its outputs.

## Acknowledgements

DMcK is receiving funding from the Irish Research Council's Government of Ireland Postgraduate Scholarship Program (GOIPG/2019/2798). The Irish LOFAR Consortium is supported by funding from Science Foundation Ireland and the Department of Further and Higher Education, Research, Innovation and Science. The compute servers of REALTA used for development of this software were funded by Science Foundation Ireland.